\documentclass[letterpaper]{IEEEtran}
\usepackage{url}
\usepackage[final]{graphicx}
\usepackage{graphicx}
\usepackage{amsmath}
\usepackage{wrapfig}
\usepackage{multirow}
\usepackage{colortbl}
\usepackage{color}
\usepackage{soul}
\usepackage{enumerate}
\usepackage[caption=false]{subfig}
\usepackage[table]{xcolor}
\usepackage{pifont}
\usepackage{paralist}

\usepackage{siunitx}

\usepackage{amssymb}
\usepackage{paralist}
\IEEEoverridecommandlockouts
\usepackage{graphicx}

\usepackage{textcomp}
\usepackage{gensymb}
\usepackage{stfloats}
\usepackage{tikz}
\usepackage{verbatim}
\usetikzlibrary{chains,positioning,shapes.symbols,arrows,shapes,shadows,trees}
\usepackage{standalone}
\usepackage{array}
\usepackage{cite}
\usepackage{cuted, nccmath}
\usepackage{lipsum}
\begin{document}
	\bstctlcite{IEEEexample:BSTcontrol}
	\title{Terahertz Wireless Communications in Space
	}
	
\author{
	Meltem Civas, Ozgur B. Akan
	\thanks{The authors are with the Department of Electrical and Electronics Engineering, Ko\c{c} University, Istanbul, 34450, Turkey  (email: \{mcivas16, akan\}@ku.edu.tr).}
	\thanks{Ozgur B. Akan is also with Internet of Everything (IoE) Group, Electrical Engineering Division, Department of Engineering, University of Cambridge, Cambridge, CB3 0FA, UK (email: oba21@cam.ac.uk).}
\thanks{This work was supported in part by Huawei Graduate Research Scholarship.}}
	
	\maketitle
	\begin{abstract}
		The New Space Era has increased communication traffic in space by new space missions led by public space agencies and private companies. Mars colonization is also targeted by crewed missions in the near future. Due to increasing space traffic near Earth and Mars, the bandwidth is getting congested. Moreover, the downlink performance of the current missions is not satisfactory in terms of delay and data rate. Therefore, to meet the increasing demand in space links, Terahertz band (0.1-10 THz) wireless communications are proposed in this study. In line with this, we discuss the major challenges that the realization of THz band space links pose and possible solutions. Moreover, we simulate Mars-space THz links for the case of a clear Mars atmosphere, and  a heavy dust storm to show that even in the worst conditions, a large bandwidth is available for Mars communication traffic. 
	\end{abstract}
	\begin{IEEEkeywords} 
Deep space communications, inter-satellite links, terahertz Communications
	\end{IEEEkeywords}
\section{Introduction}
\label{sec:intro}
The start of a \textit{New Space Era} has led to a paradigm shift in the space industry. An increasing number of private companies with space missions have emerged, and spacecraft are getting smaller and easily deployable with the help of enabling technologies. Moreover, the non-terrestrial networks are also recognized by the future release 17 of the 3rd Generation Partnership Project (3GPP) to be studied under the study item "Non-Terrestrial Networks". Vertical networks are expected to be integrated into the next generation 5G and beyond networks. This can lead to many novel paradigms including the Internet of Space Things (IoST) \cite{akyildiz2019internet}. Therefore, available bandwidths are getting congested. National Aeronautics and Space Administration (NASA) estimates that the growth factor of deep space communication capability should be at least 10 to meet the growing demand in the next three decades \cite{nasa}. Therefore, high data rate communication technologies are required. Free space optical (FSO) communication has been envisioned for space applications. However, optical communication highly depends on atmospheric conditions such as fog, cloud, and haze where optical links experience strong attenuation \cite{kaushal2016optical}. Moreover, FSO is costly, and beam alignment problems pose several additional challenges. Terahertz band (0.1-10 THz) wireless communications, which can enable high data rates on the order of Terabits per second, is an alternative to FSO.

\begin{figure}[!t]
	\includegraphics[width = \columnwidth]{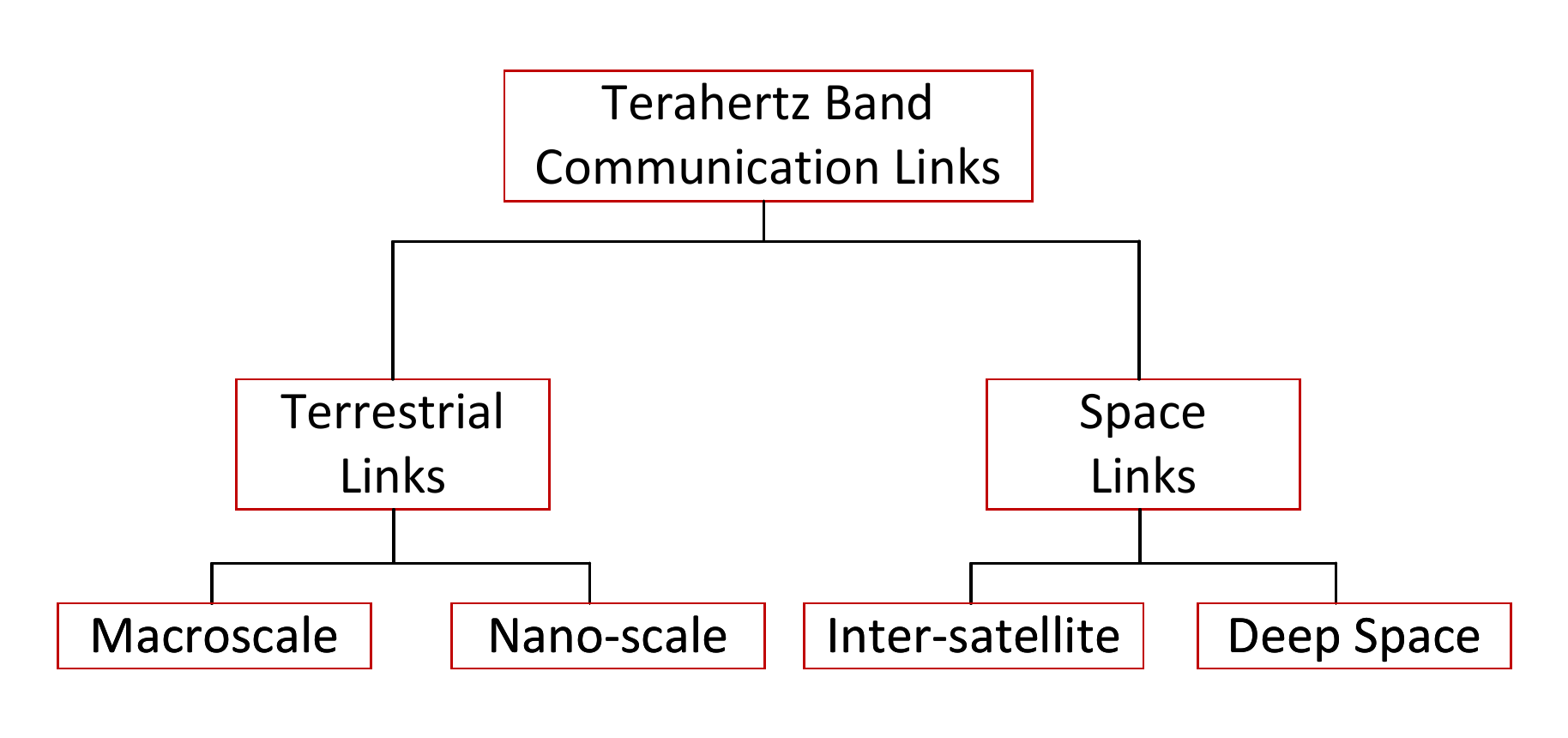}
	\caption{Classification of THz links \cite{civasterahertz}.}\label{fig:Class} 
\end{figure}

The advantages of FSO such as large bandwidth, high data rate, high security, no spectrum licensing are of growing interest.
NASA's demonstration in 2013 called Lunar Laser Communication Demonstration proved that high downlink and uplink transmission rates from lunar orbit to Earth are possible via optical links and NASA will further conduct Deep Space Optical Communications demonstration as a part of Psyche mission to show that higher data rates, 10-100 times of the current state of the art, are attainable via optical wireless communications \cite{mars_optical}. The European Data Relay Satellite System (EDRS) also employs optical inter-satellite links, which can reach a data rate of 1.8 Gbps for a transmission distance up to 45000 km \cite{esa_optical}. Apart from institutional projects, the companies such as SpaceX and Google aim to use optical communications for space-to-space or air-to-air links of their projects under development. Despite its advantages, FSO communications have still several challenges. Regarding space links, due to high transmission distances Effective Isotropic Radiated Power (EIRP) must be high. This can be enabled by large aperture transmitting optics, which result in narrow-beam divergence ($\propto \lambda/D,$ where $\lambda$ is the wavelength and $D$ is the diameter of aperture). Thus, a deviation of the optical beam from the target results in an outage. Considering the same transmitting antenna size, THz systems have looser beam pointing requirements compared to FSO links, since we consider larger wavelengths.

FSO communications are also highly affected by atmospheric conditions such as clouds, haze, fog, and atmospheric turbulence. For instance, when visibility is less than 50 m when dense fog exists, attenuation can be as high as 350 dB/km, which limits transmission  \cite{kaushal2016optical}. In case of fog with the same visibility, there are several windows in the THz band where the attenuation is below 100 dB/km \cite{schneider2012link}. Compared to FSO, THz communications are more robust to weather effects in general.
Although the attenuation in THz links for long-distance communication is still a challenge to be addressed, there are promising solutions such as large THz antenna arrays enabling extremely high gains. 
Moreover, THz waves are not affected by turbulence-induced scintillation, which is observed as intensity fluctuations at the signal, as much as FSO links \cite{han2019terahertz,kaushal2016optical}.

\begin{figure*}[!t]
	\centering
	\includegraphics[scale = 0.6]{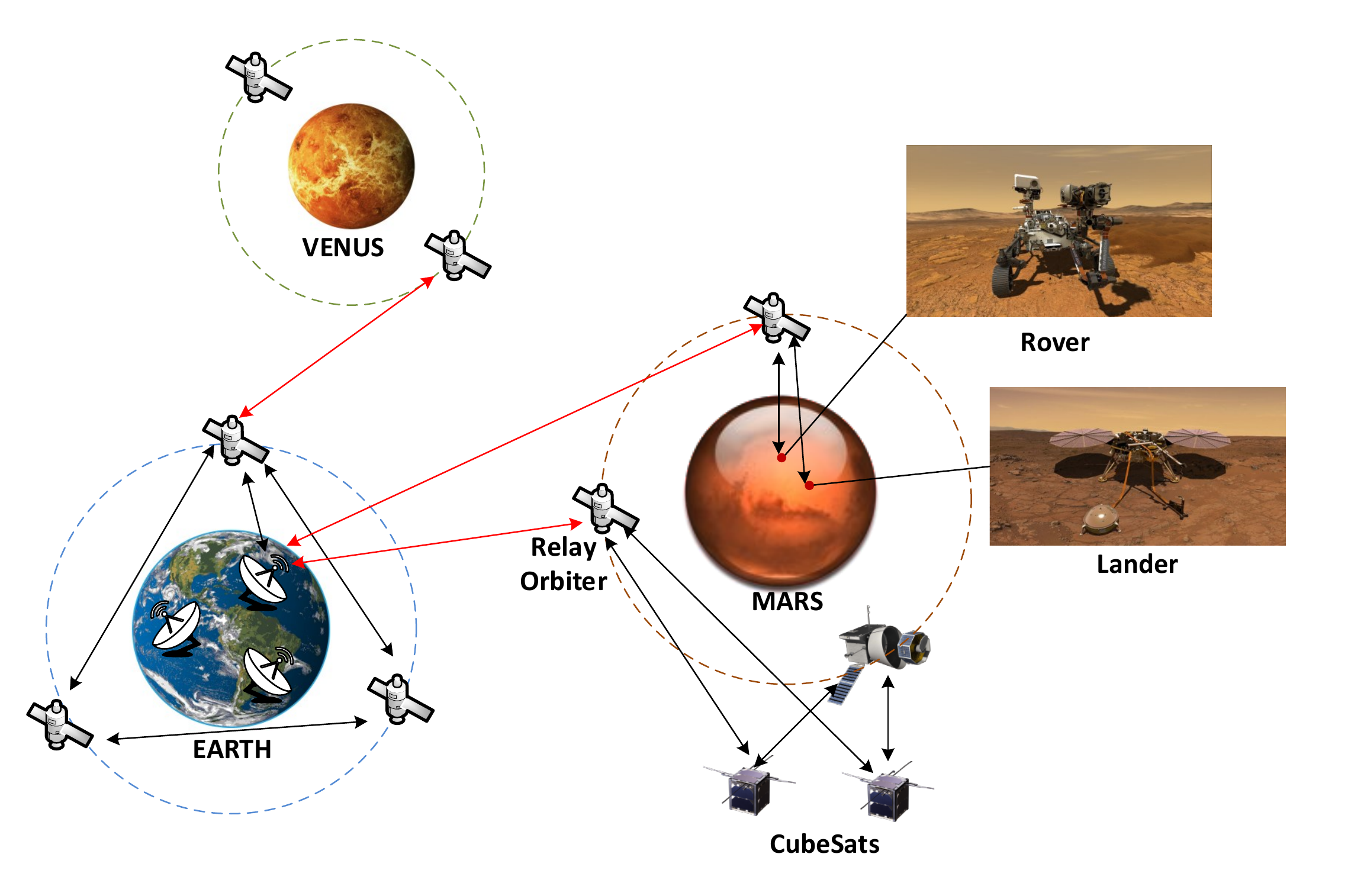}
	\caption{Interplanetary communications: Black arrows (THz links), red arrows (FSO links).}\label{<figure10>} 
\end{figure*}

THz band communication links can be classified into two as terrestrial and non-terrestrial networks as shown in Fig. \ref{fig:Class}. Terrestrial networks comprise macroscale and nano-scale links. In this study, we consider non-terrestrial network components ground/space and space links consisting of inter-satellite and deep space communication links. We aim to identify the challenges related to THz space links and discuss the possible solutions. The realization of THz space links poses several challenges. The spreading loss due to the expansion of propagating electromagnetic waves is increasing drastically with the frequency. This limits the communication distance to few meters on Earth due to immature THz source technology, which can enable transmit power on the order of milliwatts. Moreover, strong molecular absorption results in high atmospheric attenuation in Earth-to-space links; thus, limits the utilization of the high THz band. Artificial satellites occupy higher atmospheric layers of Earth or deep space where the air molecules are scarce or none. Therefore, THz inter-satellite links do not experience significant atmospheric attenuation.  
Regarding another terrestrial planet Mars, atmospheric attenuation is expected to be low compared to Earth because water molecules, which are the primary source of atmospheric attenuation, are scarce in the Mars atmosphere. These create an opportunity of utilizing the high THz band, consequently providing high data rates. In line with this, later we simulate the transmittance of Mars's atmosphere in clear and dusty atmospheric conditions using an accurate radiative transfer tool called Planetary Spectrum Generator (PSG) to show the availability of a large bandwidth for Mars communication.

The rest of the paper is organized as follows. In Section \ref{sec2}, we describe the applications of THz space links. In Section \ref{sec3}, we discuss the challenges THz band communications encounter, and then in Section \ref{sec4}, we simulate zenith transmittance of Mars atmosphere. In Section \ref{sec5}, conclusions are stated.

\section{Applications}
\label{sec2}
Utilizing THz space links can pave the way for novel applications, some of which are discussed as follows. 

\subsection{Earth observation}
Earth observation using artificial satellites began with the launch of Sputnik 1 by the former Soviet Union in 1957. Since then, many Earth observation satellites have been launched. Most of these satellites occupy Low Earth Orbits (LEO) and transmit a large amount of data to Earth daily. Recently, an increasing number of LEO satellites are being launched so that the bandwidth used is getting congested. To reduce the transmission delays and support the transmission of a large amount of sensing data to Earth, technologies supporting high data rates are required.
FSO communications, providing connectivity within a few kilometers using laser beams, have been proposed as a viable solution \cite{light_way}.  For instance, EDRS employs FSO communications between LEO satellites collecting Earth observation data and GEO satellite relaying data to Earth. However, with the start of a New Space Era, satellites are getting miniaturized, and deploying many small satellites, e.g., CubeSats, is preferred \cite{kodheli2020satellite}. The power and size requirements of FSO systems far exceed the limitations of cube/micro/nano-satellites. 
On the other hand, building THz transceivers with a large number of antenna arrays, i.e., phased Multiple-Input and Multiple-Output (MIMO) arrays, in a small footprint, is possible thanks to novel materials such as graphene \cite{civas2021universal}. Thus, for long-distance and high-data-rate near-Earth transmission, THz communications can be leveraged in the future. Large THz arrays are also advantages over FSO links in terms of beam-alignment, e.g.,  they can provide automatic alignment by their scanning ability \cite{sengupta2018terahertz}. 
However, there exist issues to realize such THz transceivers. The main impediments include the lack of practical THz signal sources and detectors, implementation and optimization of antenna arrays \cite{sengupta2018terahertz}. 

\subsection{Interplanetary communications by hybrid THz/FSO links}
Current state-of-the-art communication technologies used are not able to support high data rate interplanetary communications as a part of space information networks, which can result in numerous applications including space observation, Internet of Things (IoT), and maritime monitoring.  
To illustrate, Mars Reconnaissance Orbiter employs X-band (8-12 GHz) and Ka-band (26.5-40 GHz) to communicate with the NASA Deep Space Network, which comprises deep space communication facilities for commanding and tracking spacecraft. The data rate is between 0.5 and 4 megabits per second \cite{mars_orbiter}. The services such as live video feeding, high-resolution scientific data streaming, virtual reality for controlling rovers and other machines, and real-time data transmission will require much higher data rates.  Although latency is still an issue, these services can be enabled by hybrid THz/FSO links as a part of space information networks. In case of an outage due to beam pointing errors or strong atmospheric attenuation, THz links can be used as a backup \cite{khalighi2014survey} since THz links are more robust to weather conditions and pointing errors. 

THz communication capability can be integrated into FSO communication architecture to support high data rates and can lead to novel applications such as space exploration. For instance, Mars rovers communicate with Earth through relay orbiters since the availability of orbiters is much longer than that of rovers. A Mars rover can communicate with a relay orbiter through THz links and the relay orbiter relays messages using FSO links because THz communications are more robust compared to FSO in ground/space links. CubeSats are being used in interplanetary missions during mission-critical events, which are exemplified by Mars Cube One (MarCO). CubeSats can be equipped with
THz transceivers in the future so that they can communicate with the landing spacecraft using low latency THz links and, to relay information to Earth using FSO links. 

\begin{figure*}[t!]
	\centering
	\subfloat[][Altitude $=$ 0 km.]{\includegraphics[width=0.33\textwidth]{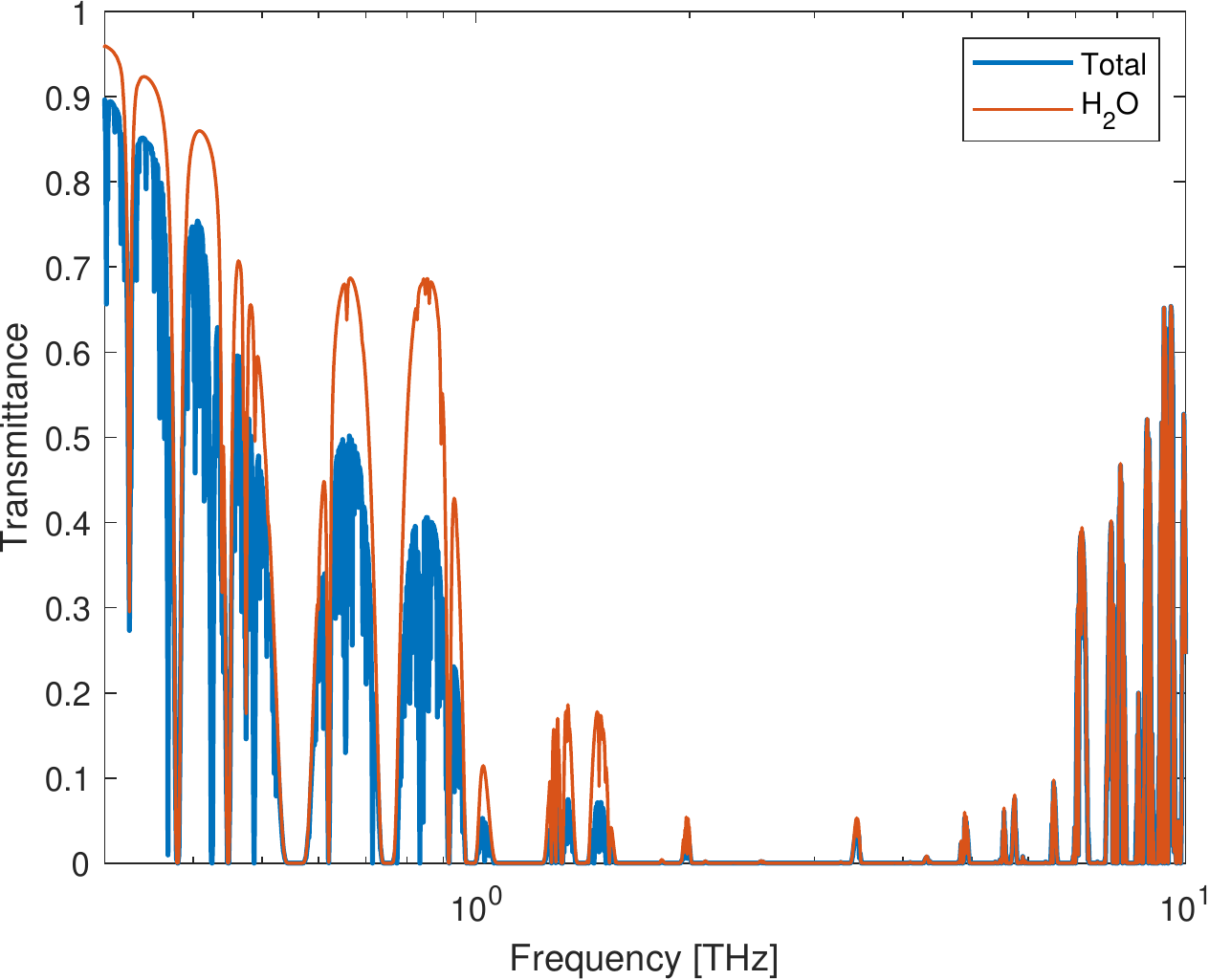}\label{<figure4>}}
	\subfloat[][Altitude $=$16km.]{\includegraphics[width=0.33\textwidth]{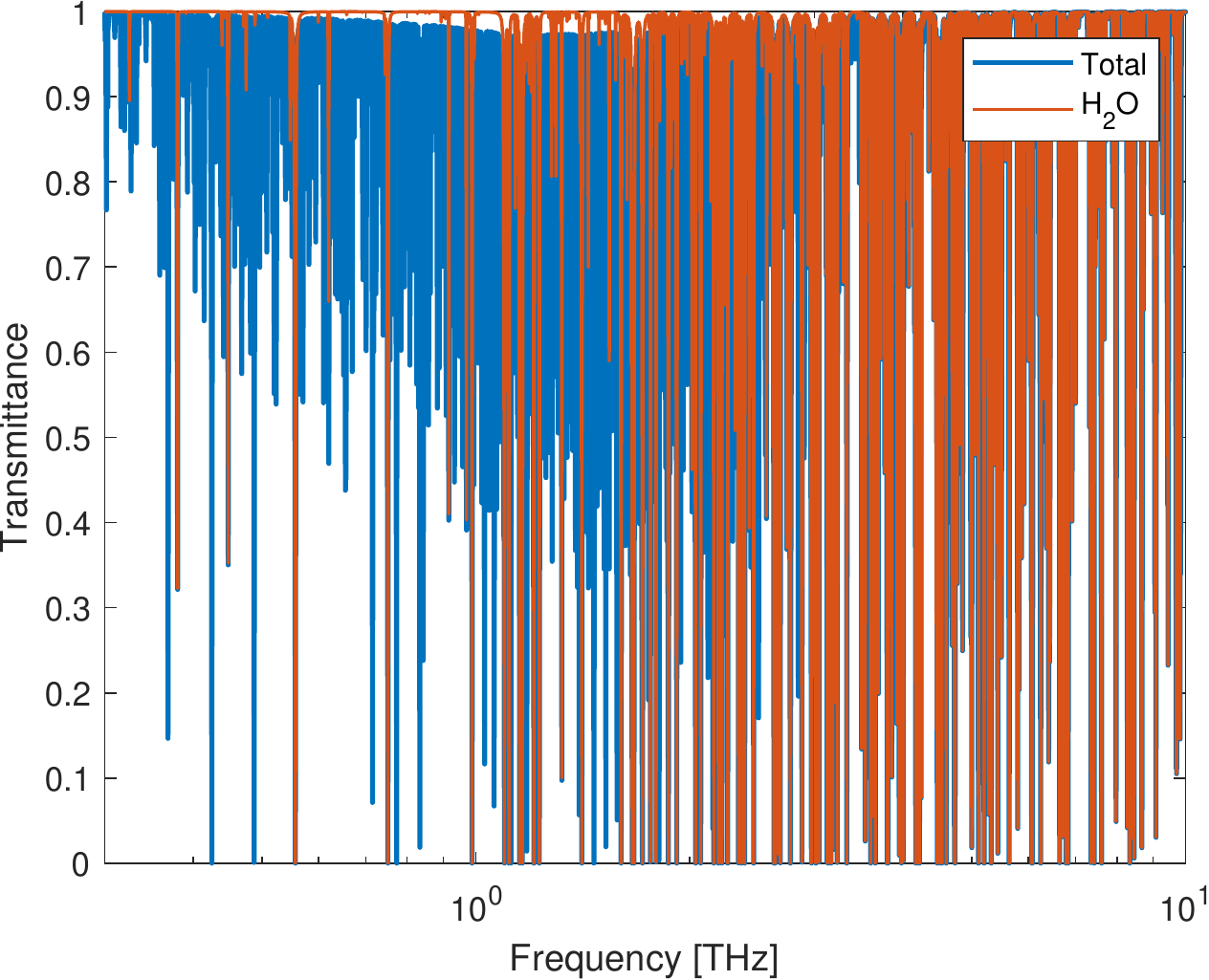}\label{<figure5>}}  
	\subfloat[][Altitude $=$35km.]{\includegraphics[width=0.33\textwidth]{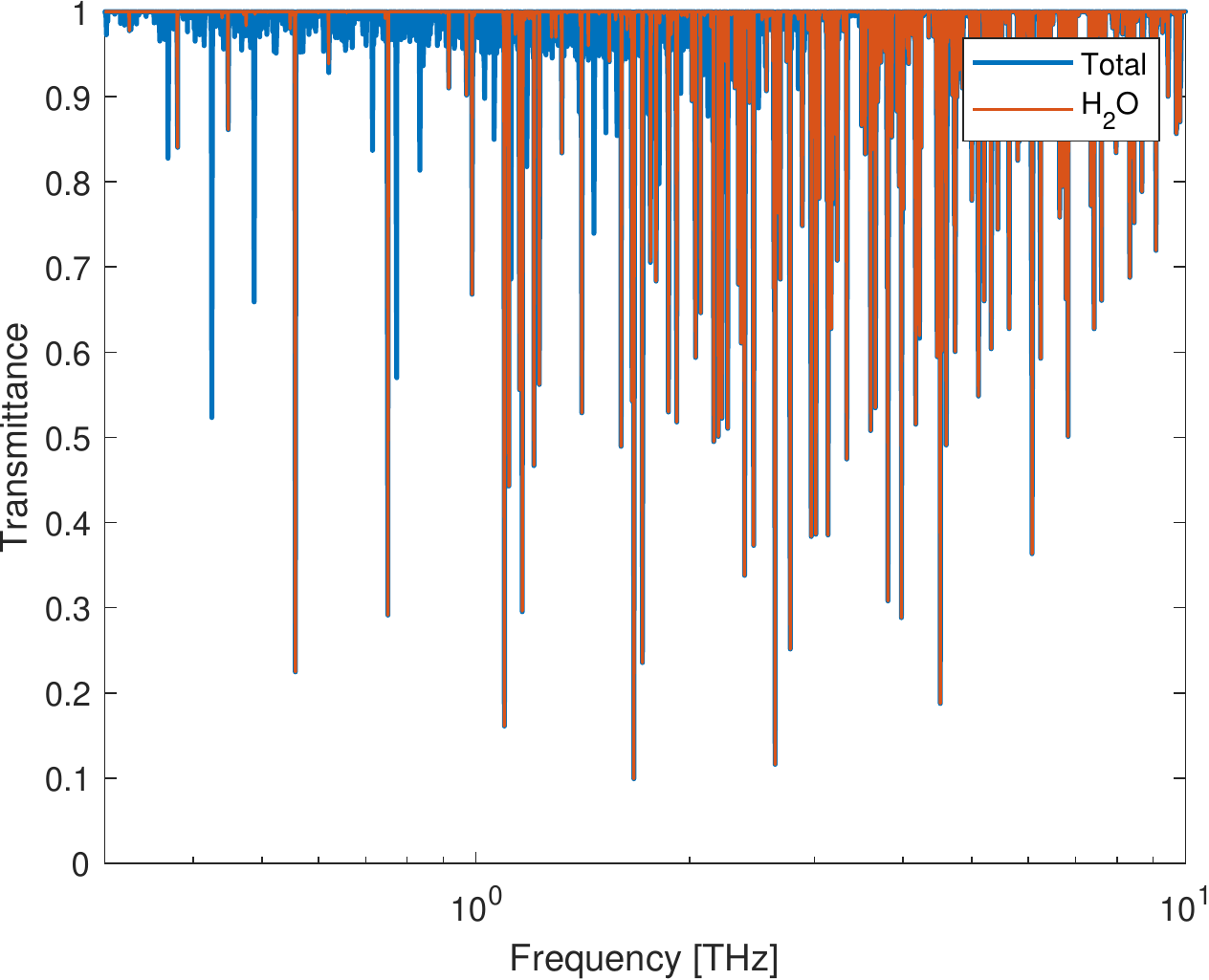}\label{<figure6>}} 
	\caption{The transmittance of Earth-space links at several altitudes from one of the driest locations of the Earth, i.e., ALMA, denoting the fraction of EM radiation after experiencing molecular absorption. The zenith angle is  $35^{\circ}$.}
	\label{fig:dry}
\end{figure*}
\section{Challenges and Solutions}
\label{sec3}
In this section, we discuss the several challenges THz space links encounter. These challenges include molecular absorption and spreading loss, interference to passive services. We also discuss the potential directions to address these challenges. 
\subsection{Molecular absorption loss}
Molecular absorption loss, which occurs when the part of wave energy is transformed into molecular energy due to the vibration of molecules, is one of the main impediments affecting Earth-space THz links. Water vapor molecules, which are scarce in the atmosphere of Mars, are the primary sources of molecular absorption on Earth in THz frequencies \cite{suen2014global}. 
Inter-satellite THz links among Low Earth Orbits (LEO), Medium Earth Orbits (MEO), and Geosynchronous Orbit (GEO) satellites are not affected by molecular absorption loss because they operate at the altitudes where water vapor is almost none. To combat the high atmospheric attenuation on Earth, several approaches have been proposed. In the following section we give an overview of these approaches. 

\subsubsection{Dry ground sites}
Atmospheric Precipitable Water Vapor (PWV) is the primary cause of strong atmospheric attenuation at THz frequencies. Considering ground-based telescopes of THz radio astronomy such as Atacama Large Millimeter/Submillimeter Array (ALMA) located at a high and dry plateau of Chile and Combined Array for Research in Millimeter-Wave Astronomy (CARMA) which was operating in the United States, Suen \textit{et al.} identified the locations on Earth with lowest water vapor \cite{suen2014global}. Dry sites for radio astronomy as well as satellite communications include Antarctica, Greenland, the Atacama Desert, and the Tibetan Plateau. Numerous dry sites that can provide acceptable performance have been identified in the United States and Europe as well.   
Suen \textit{et al.} further investigated the performance of a THz ground to geostationary satellite links \cite{suen2015modeling}. They have shown that utilizing radio astronomy platforms with large aperture antenna arrays, which are located at dry sites of Earth, for satellite communications 1 terabit/second link performance can be exceeded in clear atmosphere conditions \cite{suen2015modeling}. In \cite{zhen2018link}, the authors show that 1 terabit/second is attainable in the low THz band for ground/satellite links utilizing massive antenna arrays and establishing the ground stations at Tanggula, Tibet, where PWV is very low.

\subsubsection{High and low altitude platforms}
To combat high atmospheric attenuation in ground/satellite THz links, placing transceivers on airborne platforms has been proposed in \cite{suen2015modeling,suen2016terabit}. High and low altitude platforms such as aerostats, aircraft, and high altitude balloons can be employed for transceiver placement.
These platforms need a lower aperture diameter compared to ground-based transceivers and can also offer performances comparable to ground-based platforms with large apertures since they operate at the altitudes where water vapor density is low \cite{suen2015modeling}. In Fig. \ref{fig:dry}, Earth-space link transmittances for various altitude levels in a dry location of Chile, which are simulated in the Planetary Spectrum Generator (PSG) \cite{villanueva2018planetary}, is depicted. According to this, numerous bands, which are not feasible to use at sea level, are available for use at high altitudes. 
\subsubsection{Hybrid ground/satellite links}
Placing ground stations only at dry locations can limit the potential of THz communication. Akyildiz \textit{et al.} propose ground-satellite links enabled by microwave (e.g., X band (8-12 GHz), Ku band (12-18 GHz, Ka-band (26.5-40 GHz)), and mm-Wave/THz bands for small satellites called CubeSats in \cite{akyildiz2019new}. If a ground/satellite link is not suitable for transmission at THz frequencies, mmWave and microwave bands can be utilized. The idea is based on sending a pilot signal to obtain the availability of the link according to some criteria (e.g., Line of Sight (LOS) and weather conditions). The analysis in \cite{akyildiz2019new} shows that even with a relatively high water vapor density, data rates on the order of tens of Gbps can be achieved.   
\begin{figure*}[h]
	\centering
	\subfloat[][Altitude $=$ 0 km.]{\includegraphics[width=0.33\textwidth]{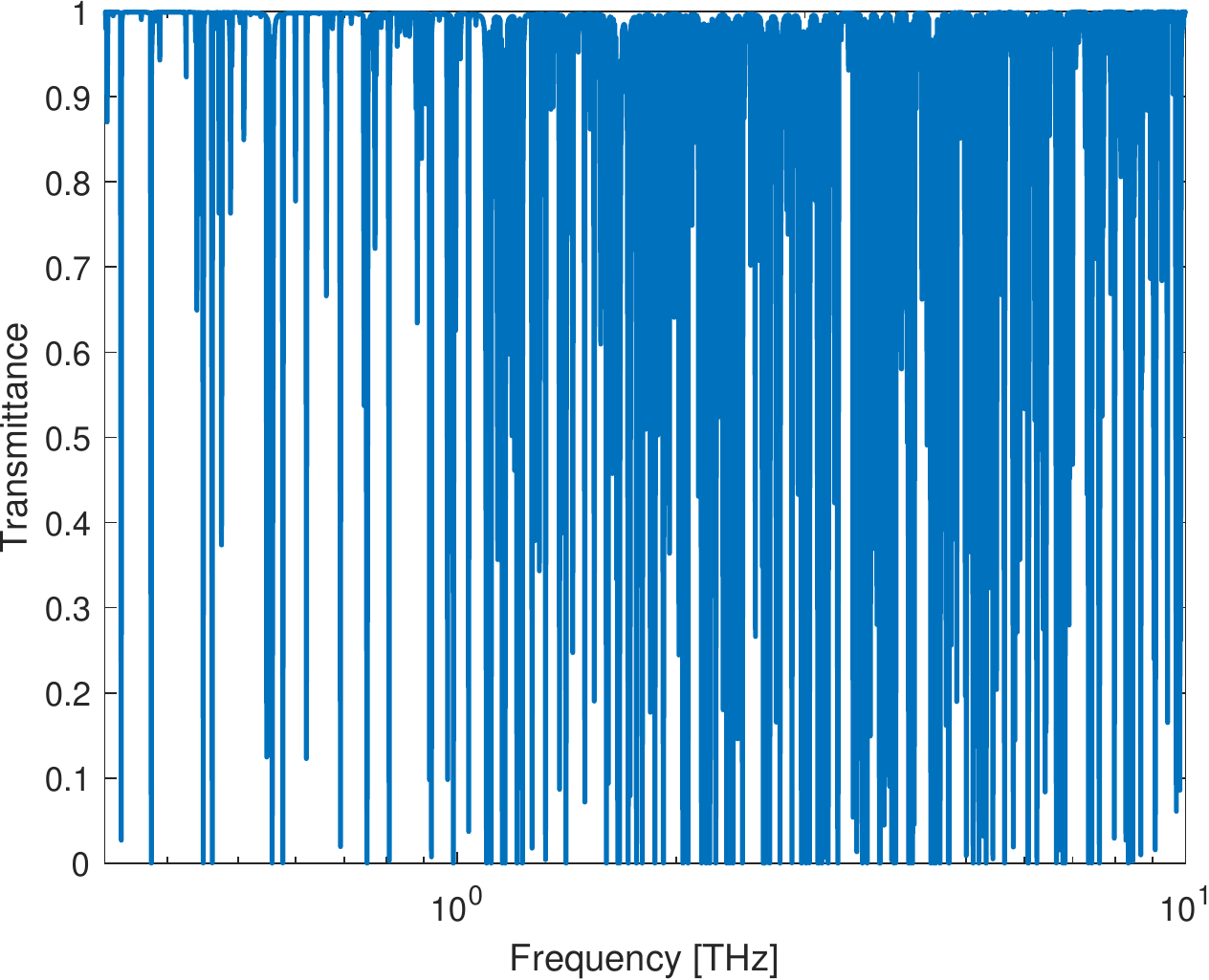}\label{<figure9>}}
	\subfloat[][Altitude $=$16km.]{\includegraphics[width=0.33\textwidth]{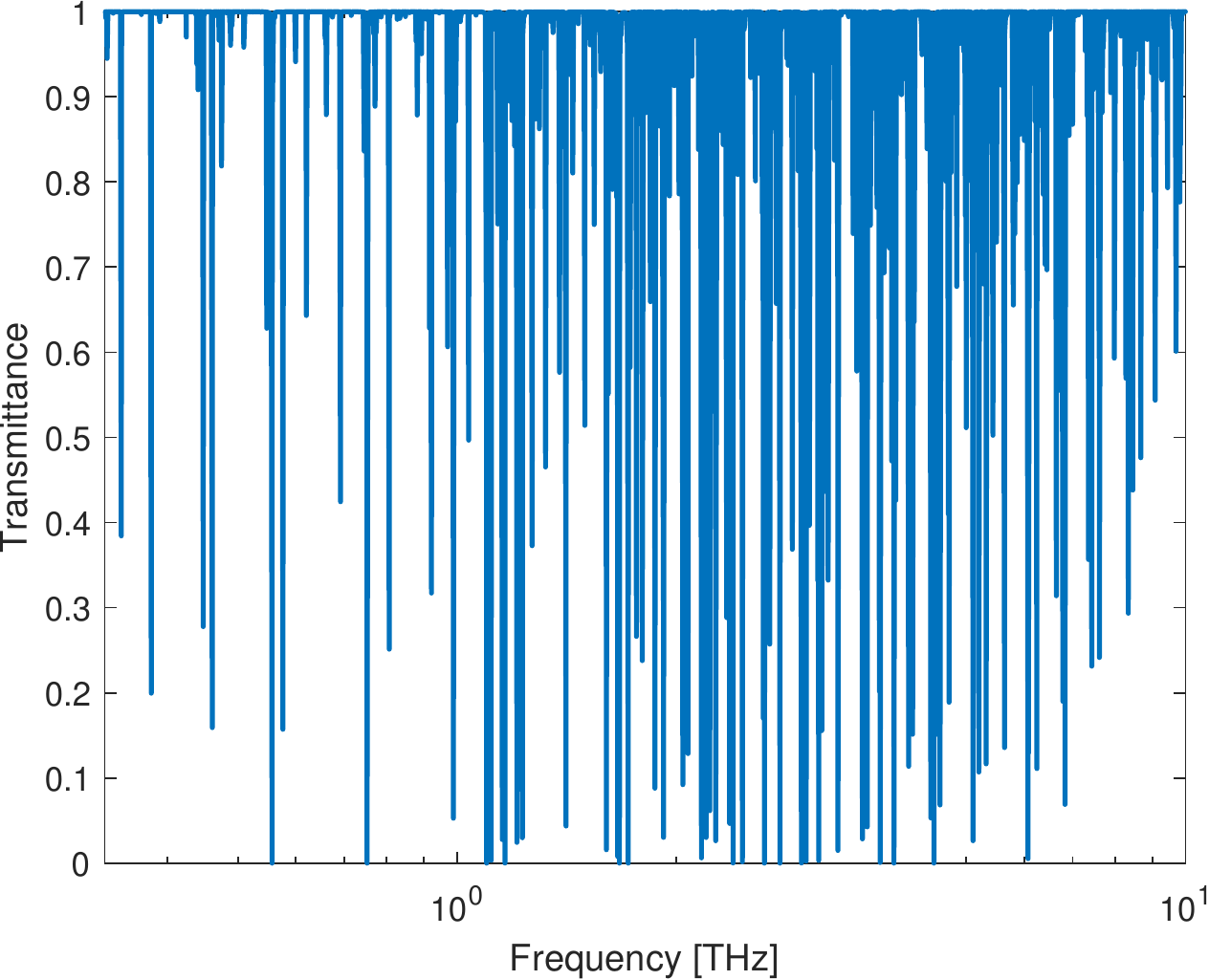}\label{<figure8>}}  
	\subfloat[][Altitude $=$30km.]{\includegraphics[width=0.33\textwidth]{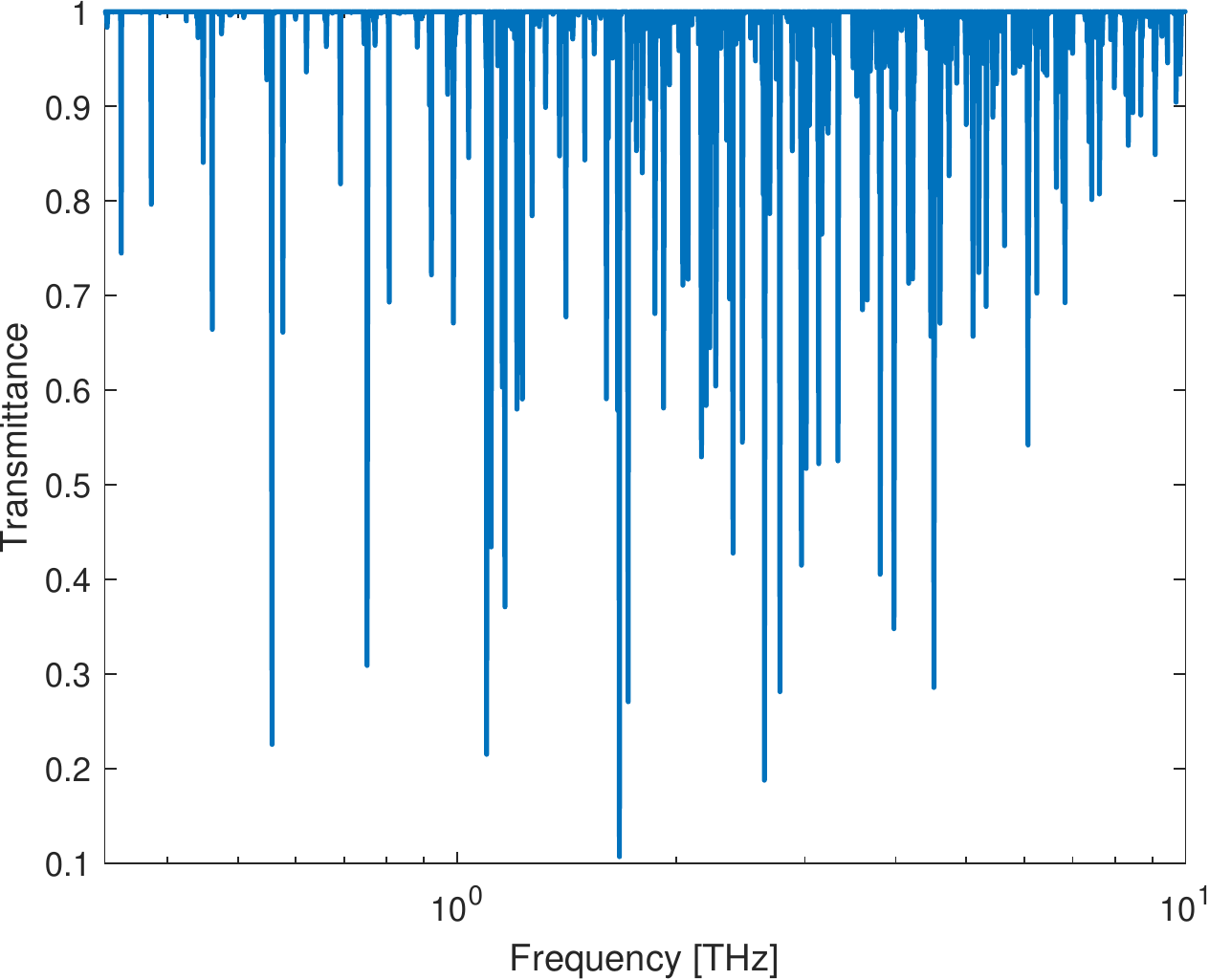}\label{<figure7>}} 
	\caption{Zenith transmittance of clear Mars atmosphere for various altitudes. }
	\label{fig:clear}
\end{figure*}

\subsection{Spreading loss} 
As an Electromagnetic (EM) wave propagates through a medium it expands and this leads to a loss called spreading loss. Spreading loss is one of the significant challenges limiting communication to short distances at THz frequencies because free space path loss increases with the frequency in a quadratic relation according to Friis' law. Regarding space links, we consider the distances at least on the order of thousands of kilometers. On the other hand, transmit power in THz frequencies is on the order of milliwatts due to immature THz source technology, which is also called the THz gap. 
Thus, high gain antennas with high directivity are required for THz space links. Several approaches for combating the problem of high propagation loss have been proposed in the literature \cite{akyildiz2018combating}. Some of these solutions apply to indoor and nano-scale communications such as intelligent surfaces controlling the behavior of an EM wave \cite{nie2019intelligent} and graphene plasmonic nano-antennas \cite{tamagnone2012reconfigurable}. In the following section, we discuss the potential solutions for THz space links.

\subsubsection{Radio astronomy optics} Large aperture THz optics, which are exemplified by ALMA comprising 54 reflector antennas with 12 meter diameter and 12 smaller antennas with a 7 meter dish diameter, are already being used by radio astronomy. In line with this, one approach is employing large aperture THz ground stations and airborne stations with smaller apertures \cite{suen2015modeling}. Large apertures can provide high gain; however, one drawback is that the construction cost increases with the diameter \cite{suen2016terabit}. 
\subsubsection{Reflect-arrays} Reflect-array antennas are being used by the systems such as satellite communications, radars and deep-space communication links \cite{hum2013reconfigurable}. 
Traditional aperture antennas can provide high gain but they are not as electronically flexible as phased arrays. However, the implementation cost of phased arrays is high. Reflect-arrays offering high gain, low cost, ease of manufacturing as well as electronic flexibility are a compromise between aperture antennas and phased arrays. To illustrate, the MarCO spacecraft communicated with Earth at a distance of 160 million km on X-band via a high-gain reflect-array antenna with small volume \cite{hodges2017deployable}. Operating frequencies of reflect-array antennas are now shifting towards THz frequencies \cite{carrasco2013reflectarray}. Thus, they offer the potential to be employed in THz space links. On the other hand, enabling technologies need to be studied because RF and MEMS technologies such as semiconductor diodes and MEMS lumped elements do not apply to the THz band due to loss and size constraints \cite{correas2017graphene}. 

\subsubsection{Ultra-massive MIMO}
The concept of Ultra-Massive Multiple Input Multiple Output (UM-MIMO) has been introduced in \cite{akyildiz2016realizing} for combating the distance problem in THz communications. According to this concept, using novel materials such as graphene to build antennas with a number of antenna elements in a small footprint is possible. Utilizing both space and frequency, the coverage range can be increased. However, the realization of this concept poses several challenges \cite{akyildiz2016realizing}. The performance of UM-MIMO depends on the THz channel; thus, accurate THz channel models are required. Moreover, to control arrays, dynamic beam-forming algorithms are needed \cite{akyildiz2019new}. 
Moreover, the performance of THz MIMO links can also be affected by the frequency-dependent diffraction of THz waves, which arises from the divergence of THz beams from their modulation side-bands. This results in degraded bit-error-rate performance due to the detection of unwanted spectrum information. Thus, novel detection and demodulation methods are required in this direction \cite{ma2017frequency}.

\subsection{Spectrum sharing}  
Spectrum beyond 275 GHz is not largely regulated in the Radio Regulations (RR). Footnote 5.565 of RR identifies numerous frequency bands in the range from 275 GHz to 1000 GHz that are used by passive services, namely Radio Astronomy Services (RAS), Earth Exploration Satellite Services (EESS), and Space Research Service, and states that the activity of these services must be protected from harmful interference of active services until the frequency allocation is established \cite{regulations2016articles}. Active and passive services will coexist on the spectrum. Current spectrum sharing studies, which aim to identify the bands where the coexistence of active and passive services is possible, mainly focus on the interference to EESS \cite{kurner2019regulatory} because RAS telescopes are located in high dry mountains. However, coexistence studies should cover both RAS and EESS because both can be affected by THz ground/satellite and inter-satellite links.

\section{Terahertz Mars-Space Links}
\label{sec4}
Preparing Mars for human exploration is one of the targets of current missions on Mars. Thus, communication among human explorers, remote-controlled vehicles, space instruments, or any other space entities will be an essential part of Mars missions soon. 
Water vapor molecules and oxygen are scarce in the Mars atmosphere. Thus, the effect of molecular absorption is less compared to Earth. However, 
Mars's atmosphere can pose other challenges due to scattering aerosols such as seasonal dust storms, limiting reliable communication.

\begin{figure}
\includegraphics[width = \columnwidth]{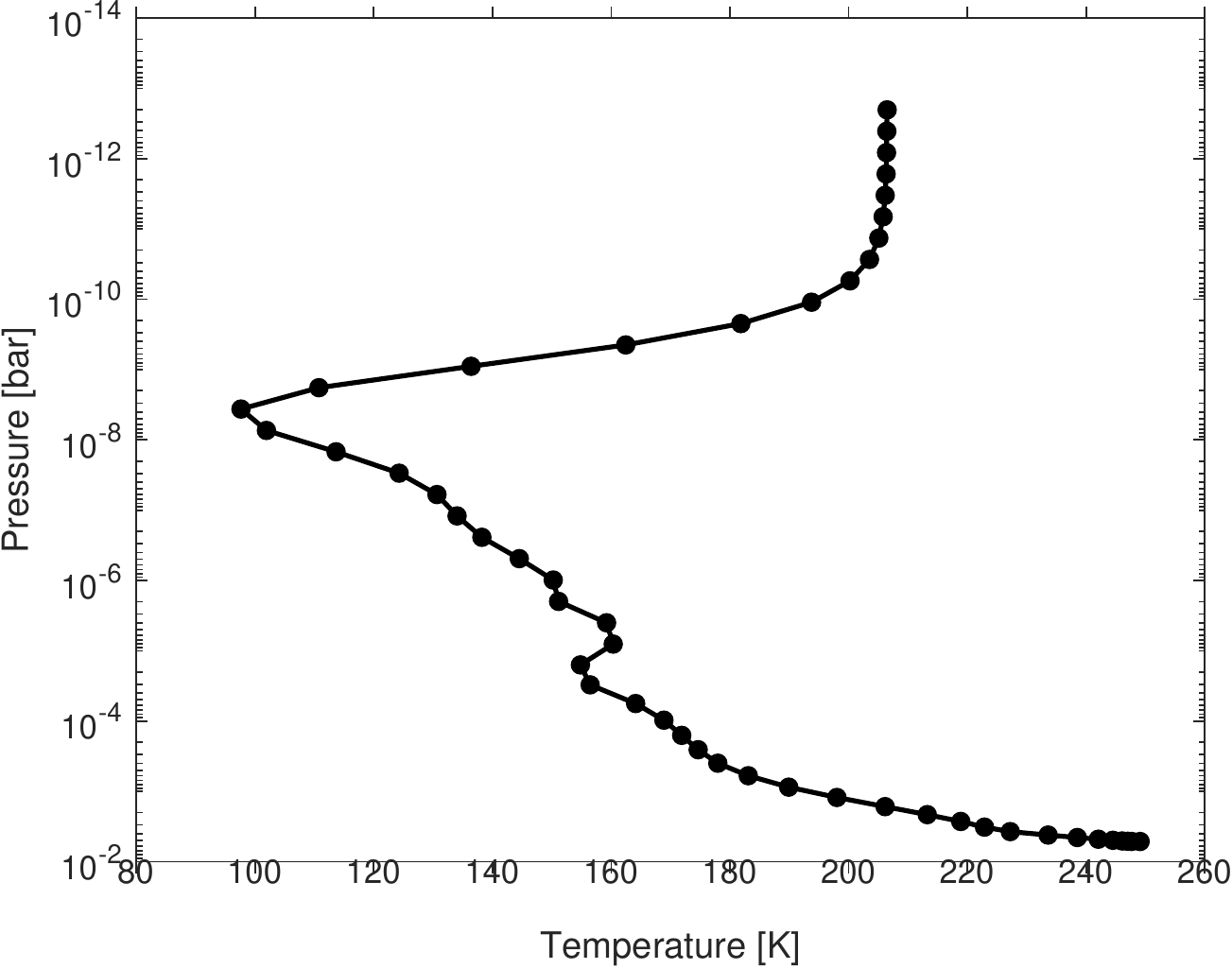}
\caption{Pressure vs temperature.}\label{fig:PT} 
\end{figure}

\begin{table}
\caption{Vertical profile of Mars atmosphere}
\label{tab:parameters1}
\begin{tabular}{|l| l|l|}	
\hline
\textbf{Gas}&  \textbf{Symbol}  & \textbf{Composition}  \\ \hline \hline
\emph{Carbon Dioxide} & \emph{$CO_2$} & $95.717 \%$ \\ 
\emph{Nitrogen} & \emph{$N_2$} & $1.991 \%$ \\
\emph{Oxygen} & \emph{$O_2$} & $0.152\%$ \\  
\emph{Carbon Monoxide} & \emph{$CO$} & $818.452$ ppm \\ 
\emph{Water Vapor} & \emph{$H_2O$} & $194.232$ ppm\\  
\emph{Ozone} & \emph{$O_3$} & $4.750$ ppb\\ 
\emph{Column} & - & 1.947e+27 m-2\\
\emph{Col mass} & - & 1.408e+2 kg/m2\\
\hline
\end{tabular}
\end{table}
\begin{figure*}[t]
\centering
\subfloat[][Altitude $=$ 0 km.]{\includegraphics[width=0.33\textwidth]{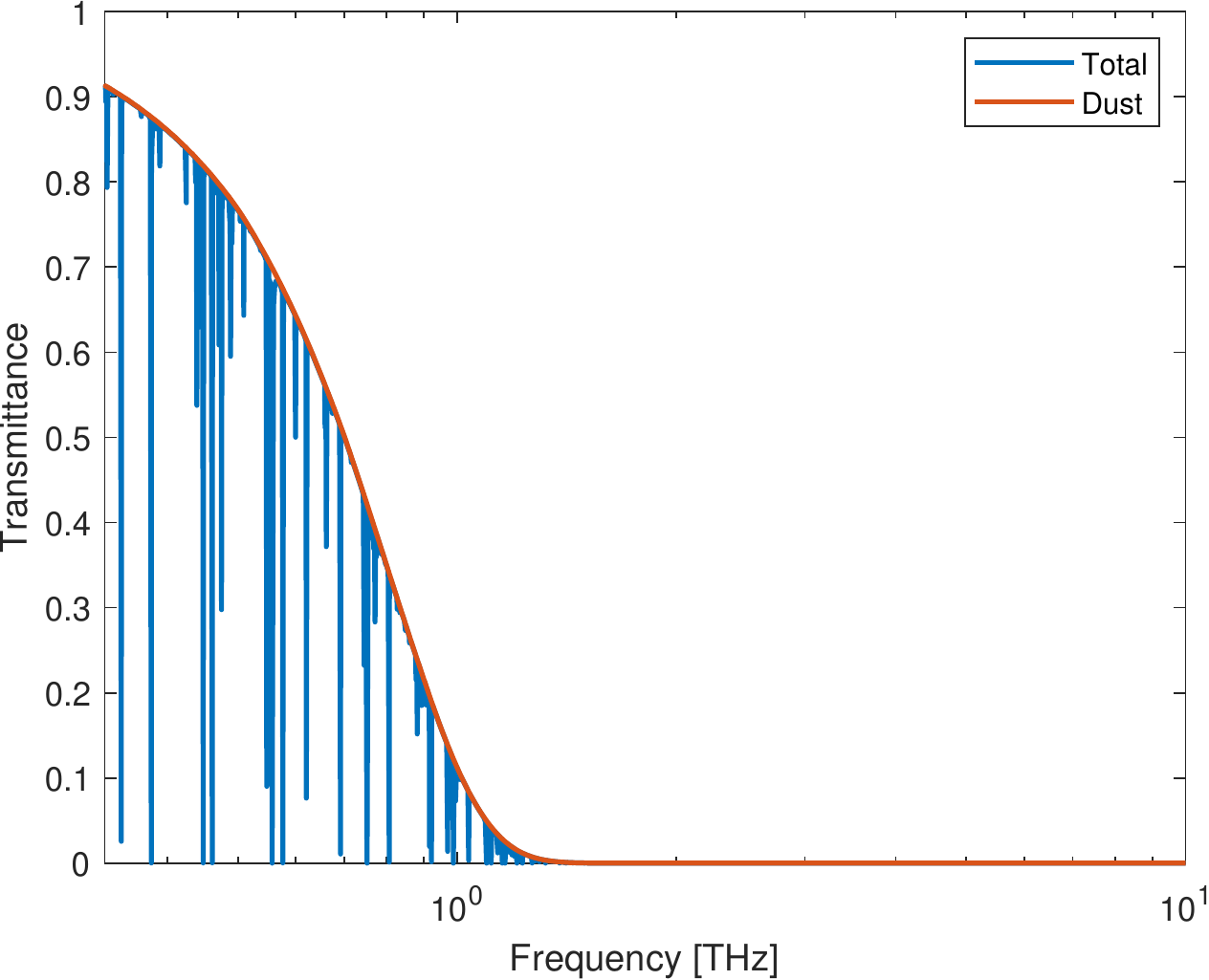}\label{<figure1>}}
\subfloat[][Altitude $=$16km.]{\includegraphics[width=0.33\textwidth]{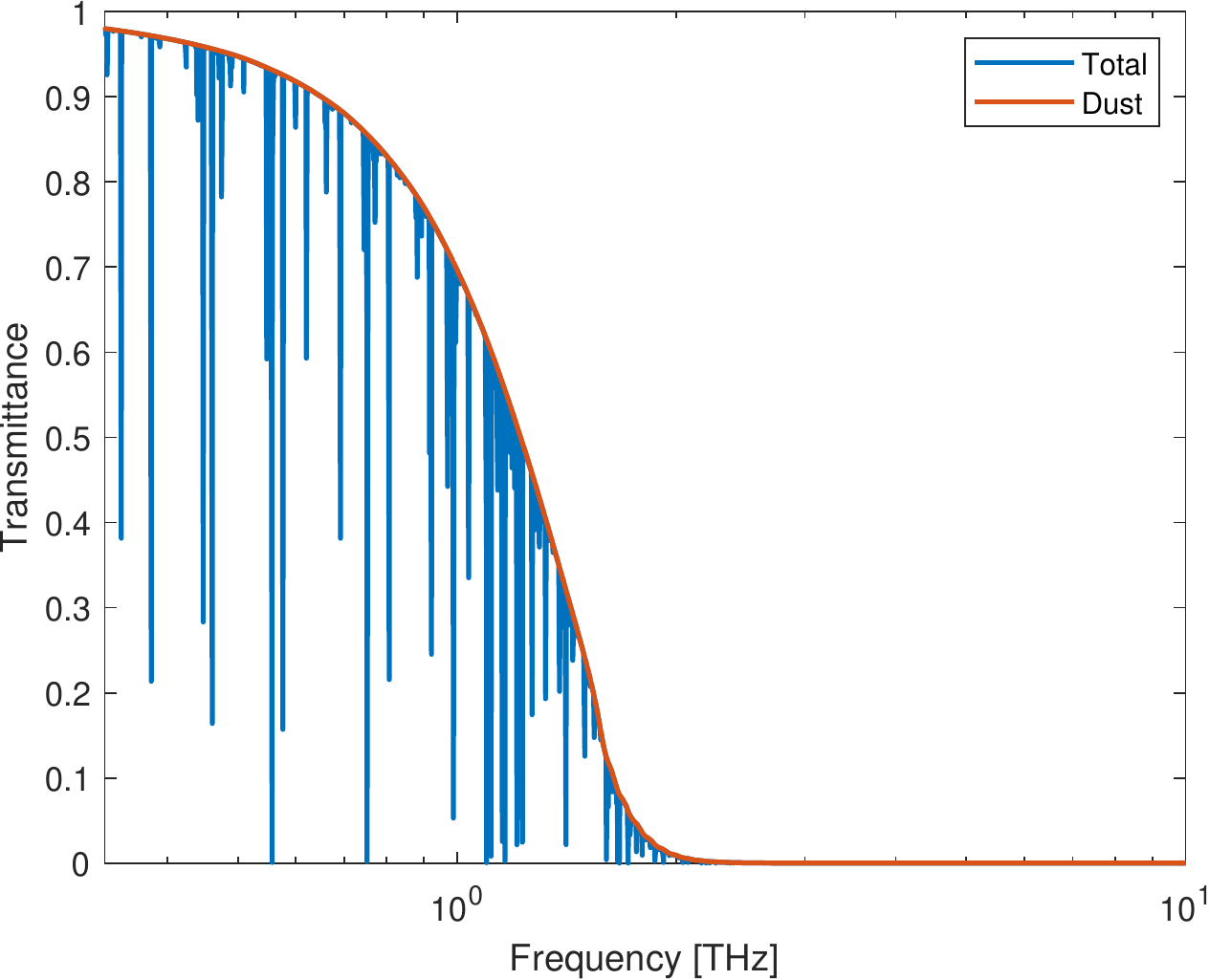}\label{<figure2>}}  
\subfloat[][Altitude $=$30km.]{\includegraphics[width=0.33\textwidth]{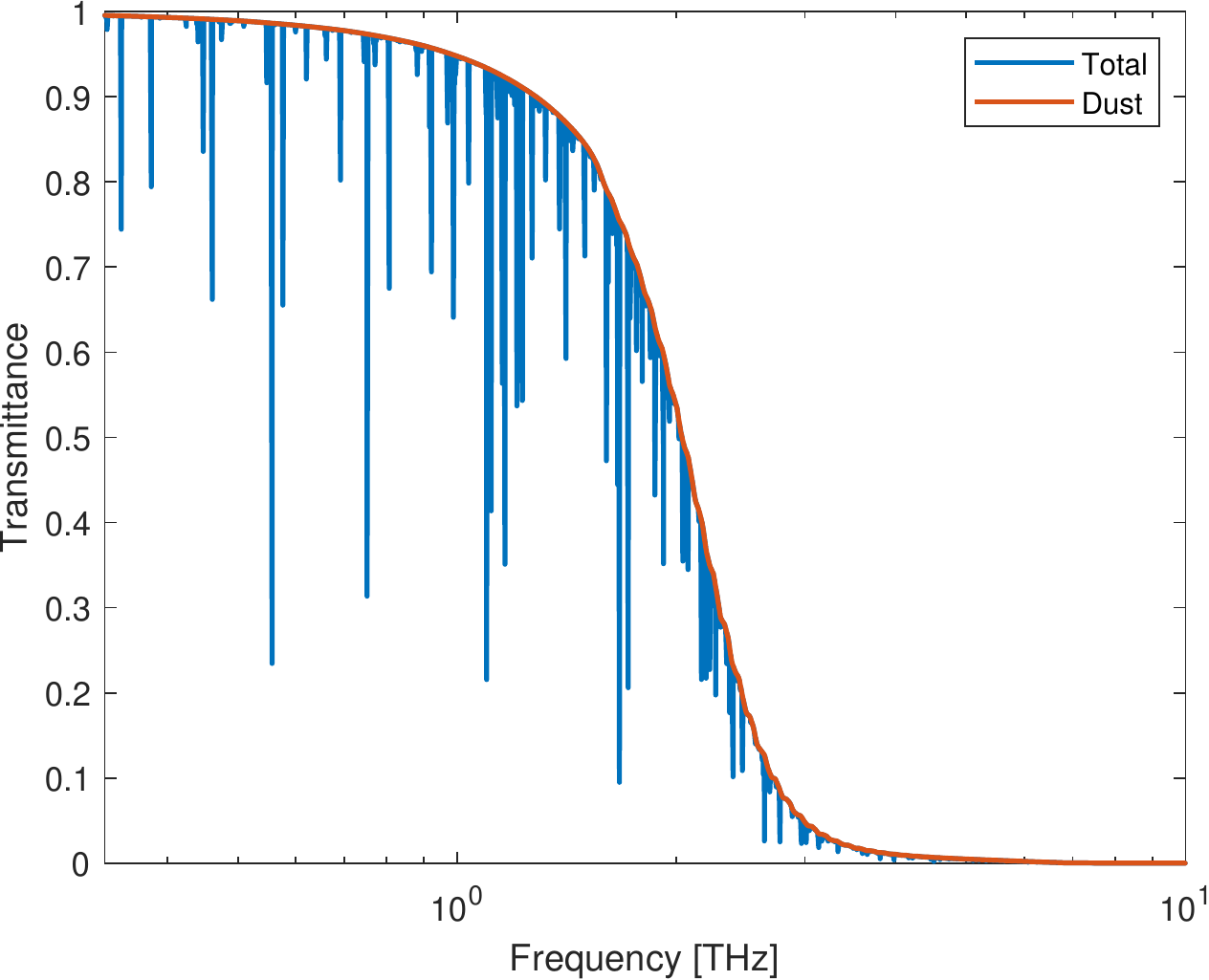}\label{<figure3>}} 
\caption{Zenith transmittance of Mars atmosphere during a local dust storm for various altitudes ( $r_{eff} = 1.5\mu m$).}
\label{fig:dust}
\end{figure*}

\subsection{The Planetary Spectrum Generator}
The PSG is an online radiative-transfer suite, which can generate planetary spectra of planets and other planetary objects \cite{villanueva2018planetary}. PSG uses several radiative transfers and scattering models, and databases including spectroscopic (e.g., high-resolution transmission database (HITRAN)) and climatological databases (e.g., Mars Climate Database (MCD)). 
We use the module of PSG extracting atmospheric profile MCD. The model of the Mars atmosphere comprises $49$ layers up to $257.90$ km. PSG further extracts information of pressure and temperature, and profiles of atmospheric gases, scattering particle sizes from MCD \cite{villanueva2018planetary}. The pressure and temperature profile used is shown in Fig. \ref{fig:PT}. The surface temperature is 279.48 K and the surface pressure is 5.1666 mbar.
Regarding the geometry, we have used the looking-up mode, in which the zenith path is considered while integrating radiative transfer. We consider the location with the longitude of 175.5 and the latitude of -14.8 (Mars Exploration Rover A landing site). The zenith angle is $25.045$ and the azimuth angle is $297.909$ in the simulations. Mars date considered is 2018/05/07.

\subsection{Zenith transmittance}

For a clear Mars atmosphere, the zenith transmittance of a Mars-space link is shown in Fig. \ref{fig:clear}. When the altitude is 0 km, it can be observed that in the $0.3-10$ THz band, transmittance values are greater than $0.9$. There are numerous sharp decreases in the band ($1-10$ THz). At the altitude of $30$ km, the whole band is available except for sharp decreases in the transmittance, since molecular absorption is not so effective at high altitudes due to the very low abundance of molecules.

Dust storms are an important phenomenon of Mars. They can be classified as local ($<2000\mathrm{km}^2$), regional ($>2000\mathrm{km}^2$) and planet-encircling \cite{ho2002radio}. Global dust storms occur in the southern spring and summer seasons of Mars.  
Dust devils are common phenomena seen both on Earth and Mars. They inject a high amount of dust into the atmosphere. During Mars southern spring and summer $0.9$ to $2.9 \times 10^{11}$ kg/$\mathrm{km}^2$ dust devil flux is estimated in \cite{whelley2008distribution}. The amount of dust injected into the atmosphere in local and regional dust storms is also reported comparable to dust devils. Accordingly, in a local storm the abundance of dust in the atmosphere can be calculated as $0.145\times 10^{6} \mathrm{kg}/\mathrm{m}^2$ (for $2000\mathrm{km}^2$ area) for the worst case. Mean or effective radius $r_{eff}$ is another important parameter to examine the effect of dust on the scattering. We assume the abundance of dust is $0.145\times 10^{6} \mathrm{kg}/\mathrm{m}^2$ in every 49 layers of the atmosphere. 
According to observation of Mars atmosphere, $r_{eff}$ varies between $1.5\mu m$ and $1.6\mu m$ \cite{smith2008spacecraft}. Thus, we consider the effect of a local dust storm on THz-band transmission in a Mars-space link at various altitudes for $r_{eff} = 1.6 \mu$m. The results in Fig. \ref{fig:dust} show that even in a heavy dust storm, transmittance values are close to 0.9 in the low THz band when the altitude is 0, which is the worst case. When the altitude is increased, transmittance values are higher than 0.9 up to 1 THz.

\section{Conclusion}
\label{sec5}
THz space links can pave the way for services including live video feeding, high-resolution imagery, and virtual reality from space. Apart from its inherent challenges such as molecular absorption loss, THz band communications pose other challenges due to developing THz source and antenna technologies. Therefore, in this study, we discuss the major challenges of THz space links, namely molecular absorption loss, spreading loss, and interference from RAS and EESS. The possible solutions for ground/space links are locating ground stations to high and dry locations, and multiband antennas. Regarding spreading loss, large aperture antennas can be used for ground stations benefiting from radio astronomy optics. Reflect-arrays and UM-MIMO are other solutions.  

\bibliographystyle{IEEEtran}

\bibliography{References}
\end{document}